# Strengthening national capability in urban climate science - an Australian perspective

Negin Nazarian[1,2,3], Andy J Pitman[3], Mathew J Lipson[2,3], Melissa A Hart[2,4], Helen Cleugh[5], Ian Harman[6], Marcus J Thatcher[6], Annette L Hirsch[7], Giovanni Di Virgilio[8], Matthew L Riley[8], Nigel Tapper[9], Jason P Evans[2,3], Christian Jakob[2,9], Pascal Perez[10]

[1]School of Built Environment, University of New South Wales, Sydney, NSW, Australia; [2]ARC Centre of Excellence for the Weather of the 21st Century, Australia; [3]Climate Change Research Centre, University of New South Wales, Sydney, NSW, Australia; [4]Insitute for Marine and Antarctic Studies, University of Tasmania, Hobart, TAS, Australia; [5]Institute for Climate Energy and Disaster Solutions, Australian National University, Acton, ACT, Australia; [6]Commonwealth Scientific and Industrial Research Organisation Environment, Aspendale, VIC, Australia; [7]Deloitte, Canberra, ACT, Australia; [8]NSW Department of Climate Change, Energy, Environment, and Water, NSW, Australia; [9]School of Earth Atmosphere and Environment, Monash University, Melbourne, Vic, Australia; [10]Faculty of Architecture, Building and Planning, University of Melbourne, Melbourne, Vic, Australia

## Abstract

Cities are experiencing significant warming and more frequent climate extremes, increasing risks for the over 90% of Australians who live in urban areas. Cities also influence the humidity, wind speeds, turbulence and air quality. However, many of our tools for climate predictions and projections lack accurate representations of urban environments. We also lack the necessary observations and datasets to assess our climate models. In this paper, we identify critical gaps, detailing how they undermine the accuracy of climate impact and risk assessments in cities, and how they may lead to poorly designed and implemented adaptation and mitigation strategies. These gaps, and the recommendations provided to address them, were identified through consultation with key Australian experts in scientific and operational research from various research institutes, universities, two Australian Research Council Centres of Excellence, federal and state governments, and private agencies.

Our recommendations span four key areas: a) city-descriptive datasets, b) integrated observations in Australian cities, c) fit-for-purpose models, and d) a coordinated community of research and practice. There is an urgent need to develop, incorporate, and tailor models that reflect the unique characteristics of Australian urban landscapes and climate. To achieve this, we require comprehensive, nationally consistent, and high-resolution datasets that accurately represent the urban characteristics (form, fabric, and function) of our contemporary and future cities. Furthermore, we urgently need to fill the systematic gaps in the integrated network of urban climate observations for systematic evaluation. In parallel, scientific understanding of key urban processes that influence weather and climate patterns must be advanced and closely integrated with improvements in how these processes are represented in physical models. This can be accomplished by establishing a national community of research and practice that co-designs and oversees an implementation plan, integrated with existing national research infrastructure programs such as the Australian Community Climate and Earth System Simulator - National Research Infrastructure (ACCESS-NRI) and the Australian Urban Research Infrastructure Network (AURIN). Developing and enhancing this national capability will enable us to answer some of the most critical questions about the interaction between cities and climate, ultimately protecting Australia's urban populations and ensuring a resilient future for our cities.

**Plain Language Summary (150 words)**

In Australia, many of our tools for climate predictions and projections lack accurate representations of urban environments. We also lack the necessary observations and datasets to assess our climate models. This perspective explores these critical gaps, common across many cities of the world, identifying how they may undermine the accuracy of climate impact and risk assessments in cities. We then provide 18 recommendations to address them, which were identified through consultation with key Australian experts in scientific and operational research from various research institutes, universities, government, and private agencies.

Our recommendations cover city-descriptive datasets, integrated observations in Australian cities, fit-for-purpose models, and a coordinated community of research and practice. Developing and enhancing our national capability will enable us to answer some of the most critical questions about the interaction between cities and climate, ultimately protecting Australia's urban populations and ensuring a resilient future for our cities.

# I. Why Australia Needs a National Urban Climate Modelling Capability

Cities are hotspots for climate change vulnerability. In Australia, 90% of the population resides in urban areas and 87% live within 50 km of the coast[1]. Cities also play a critical role in our climate change discourse, as they are major sources of greenhouse gases[2], particularly in Australia, where seven capital cities contribute 70% share of the country's GDP[3]. To ensure that we effectively mitigate and adapt to climate change and address society's increasing vulnerability to weather and climate hazards, we need a more focused and targeted urban climate research program that enhances our predictive capabilities in Australian cities.

Modelling is our main tool for exploring how various strategies for urban policy and development might interact with weather and climate. Australia has a growing community of urban climate researchers dedicated to advancing this field, particularly focused on quantifying and addressing local-scale urban climate challenges such as heat and air quality[4–6] and their implications on urban design and human health[7,8]. In urban climate modelling, Australia has considerable expertise in many important components of urban science. These include micro- and meso-scale urban climate simulations as well as the development of urban canopy parameterisations[9–14].

However, urban climate research has been largely ignored in our global and regional climate projections and underrepresented in assessing the physical basis of climate change[15,16]. A large fraction of climate research conducted to date has necessarily used coarse-resolution global models. The coarse nature of these models historically enabled urban landscapes to be ignored. Recent advances in modelling, particularly the development of high-resolution global models and existing regional models, now make it possible, and necessary, to include urban areas explicitly. Yet this underrepresentation persists, constraining our ability to provide reliable assessments of current and future climate risks for society, the environment, and the economy.

Even when urban processes are considered, Australia's modelling efforts heavily rely on models developed internationally, particularly in the US and UK. However, these schemes are not fully evaluated for their suitability in Australian cities, where key urban features differ. For instance, Australian cities feature a sizable suburban component where modelling in-canyon vegetation is critical. The types of vegetation and their interaction with the urban water balance differ markedly from those in many Northern Hemisphere cities. In addition, patterns of anthropogenic heat release (i.e., the heat generated and released into the environment due to human activities) vary due to distinctive energy use (including the balance between heating and cooling) and transportation behaviours[17]. Our reliance on international models (and underlying datasets), without systematic evaluation tailored to Australian climates and urban characteristics, restricts our ability to address fundamental scientific questions relevant to our contemporary and future cities, especially in understanding the interactions between urbanisation and climate change that drive weather and climate extremes. Moreover, this dependence leaves Australia vulnerable to shifts in foreign policy or research priorities. Recent changes in political leadership in the US, for instance, have led to funding cuts to agencies such as the National Oceanic and Atmospheric Administration (NOAA) in the United States, which maintain key models and datasets across weather and climate fields. These cuts have had global ripple effects, including threats to Australia's ability to forecast weather[18] and disruptions to global efforts to assess public health impacts[19]. In response to similar risks, the European Union has begun actively reducing its reliance on external scientific infrastructure and data streams, seeking to build its own independent systems[20]. Similarly, to ensure continuity, sovereignty, and fit-for-purpose urban climate research and forecasting capabilities, Australia must now take deliberate steps to develop and sustain its own modelling capabilities tailored to Australian cities. Strengthening domestic capability is therefore not only a scientific necessity but a strategic imperative for national resilience.

This discussion of the importance of incorporating cities into our national climate discourse and capability is not a new topic. Nearly 15 years ago, the former Chief of the Bureau of Meteorology Research Centre and the Australian Academy of Technology and Engineering (ATSE) ran a workshop titled "Climate Change and the Urban Environment: Managing Our Urban Areas in a Changing Climate", hereafter referred to as the ATSE Review[21]. The ATSE Review highlighted many of the issues we see today: a lack of systematic urban representation in our national climate modelling framework, the necessity for a "systematic measurement program including meteorological, flux, and remote sensing data", and a limited understanding of climate thresholds that trigger vulnerabilities to climate extremes in urban areas.

However, virtually none of the recommendations outlined by the ATSE Review have been addressed, particularly regarding the development of national capabilities for urban climate modelling or the necessary observational networks. Consequently, policymakers, planners, architects, land use planners, and engineers are often lacking the knowledge, tools, and data to fully understand how climate change will play out over our cities. A key deficiency has been the failure to understand and communicate the importance of urban-resolving models and the consequences of our limited urban modelling capabilities in Australia.

It is crucial to reiterate the value and urgency of this priority, now more than ever. First and foremost, the Australian population has been steadily growing over the last three decades and is projected to reach 45.9 million people by 2071[22]. The majority of this increase is expected to occur in our seven capital cities. The importance of cities in future climates will not diminish, but rather become more significant.

Second, compared to 15 years ago, we have more evidence of the errors associated with excluding cities from climate projections. In Global Circulation Models (GCMs), the omission of urban processes leads to an underestimation of warming, with average additional warming of 4°C in cities by the end of the century[23]. At local and regional scales, these errors are even greater, due to the clear two-way interactions between cities and climate change[10]. Urbanisation can exacerbate or mitigate heatwaves[24,25], alter boundary layer circulation and the water cycle, and change patterns of precipitation over and downwind of cities[26]. Many climate projections do not account for these two-way interactions in cities, limiting their ability to assess physical climate risks in cities[27]. Without the tools to quantify the interactions between Australian cities and regional and global climate change, and better understand their processes and drivers, there remain many unknown aspects of future city climates.

There have been some positive developments that have motivated swift actions now. Over the last decade, significant advances in supercomputing and computational methods have led to higher-resolution climate modelling approaching convection-permitting scales, making the omission of urban areas increasingly problematic. While models at >50 km resolution might justify focusing on broader-scale processes, finer resolution simulations suffer from significant biases if urban areas are neglected. This is especially the case for kilometre-scale simulations that some major modelling and research centres, in Australia and abroad, are now performing[28]. However, higher resolution alone is insufficient for accurate predictions over cities. The benefits of higher resolution can be undermined by poor representation of surface-atmosphere processes (in this case, urban-relevant physics) represented and the quality of land surface (i.e., city-descriptive) data. If the underlying Land Surface Models (LSMs) or land data lack appropriate detail or realism, higher resolution may yield little benefit, or even introduce new model errors[29]. Accurate urban climate projections require not just higher resolution, but also land models and datasets that faithfully capture local urban features and processes.

The second positive developments that enable action now in Australia includes the establishment of two key national research infrastructures to support development and research within the climate and urban fields, notably the Australian Community Climate and Earth System Simulator - National Research Infrastructure (ACCESS-NRI) and the Australian Urban Research Infrastructure Network (AURIN). This progress is complemented by a growing number of Australian experts in urban climate research, as well as increased cross-institutional collaborations with multiple agencies involved via climate-focused Australian Research Council Centres of Excellence, strong university engagement, and emerging stakeholder needs around urban climate modelling and projections that can respond to climate adaptation and mitigation in cities and enhance our ability to meet business needs. While these initiatives serve multidisciplinary communities with broad interests in climate and urban studies, there is no national infrastructure that currently supports and streamlines the development, data, and knowledge exchange between the two, i.e., no national infrastructure supports urban climate research.

Lastly, it's critical to remember that while urban areas are exposed to various climate hazards, they also present significant opportunities for mitigation and adaptation. Cities, for example, can create more thermally acceptable conditions during extreme heat events through climate-sensitive design and planning[10,30]. Accurate methods to simulate and assess these opportunities are key to reducing emissions and exposing fewer people to climate hazards.

Considering the importance and urgency of this topic, we revisited the ATSE Review with a more refined scope of urban climate modelling and projections, through consultation with experts in urban climate modelling and observational analyses in Australia. Participants included representatives from different research institutes, universities, two ARC Centres of Excellence, federal and state governments, and private agencies for scientific and operational research. The consultation highlighted the most critical research questions enabled by enhanced national capabilities, particularly those related to the synergies between cities and climate and weather extremes.

## II. Building National Capability: Gaps and Recommendations for Coordinated Action

This section details the gaps and recommendations identified as part of the expert consultation for urban climate modelling in Australia. These challenges are summarised into four priority areas: 1) high-resolution, city-descriptive datasets, 2) integrated observational networks, 3) model development and assessment, and 4) research community and knowledge development. Section II.A addresses the first two, which together form the foundation of Australia's data infrastructure for urban climate research and modelling. Section II.B focuses on the development and assessment of fit-for-purpose urban climate models. Section II.C outlines the steps needed to build and sustain an active, coordinated research and practice community to advance Australia's understanding of weather–climate interactions in cities.

## II.A. Data infrastructure for urban climate research and modelling

High-resolution, nationally consistent data and observational networks are the foundation of accurate and locally relevant climate science. Australia's urban environments are complex and spatially diverse, with differences in land cover, building form, vegetation, and energy use that interact with local weather and climate processes. Capturing these dynamics requires detailed datasets on the physical characteristics of cities and robust, multi-method observation systems. Importantly, these data are valuable in their own right, providing insights into the functioning of urban climates and enabling a deeper understanding of how cities respond to, and influence, weather and climate across scales. They also underpin model development and evaluation, ensuring that urban climate models remain accurate, relevant, and able to support decision-making. This section outlines two critical components of data infrastructure: a) high-resolution city-descriptive datasets, and b) integrated urban climate observation networks.

### a. High-resolution, consistent data on urban form, fabric, and function in Australian cities

Australian cities and townships have evolved from early timber and brick structures, often roofed with corrugated iron, to modern designs that, at least to some degree, reflect the country's climatic and economic conditions. These include the widespread use of verandas in subtropical regions, multi-storey apartments in urban centres, and, increasingly, energy-efficient glazing and insulation in newer buildings in cooler climates. As a result, our urban areas have diverse land characteristics with varying surface materials, building densities and heights, and heat generation. These factors significantly influence local weather patterns and produce different vulnerabilities to climate extremes.

However, there are key gaps in how we represent urban areas in our climate predictions and projections. Current modelling approaches often oversimplify these dynamics, treating entire cities as solid concrete slabs (as in current Bureau of Meteorology forecasts) or ignoring cities altogether (as in the Community Atmosphere Biosphere Land Exchange (CABLE), the LSM supported by ACCESS-NRI). This is problematic in Australia with a high urban population, and large cities dominated by suburban layouts and significant vegetation cover, and when assessing the intra-urban variability in heat and flood risks[31].

Some land cover datasets describe urban characteristics in a class-based approach, Local Climate Zones (LCZs), whereby urban areas are classified into a limited number of typologies with a resulting loss of fidelity[32]. While LCZs offer a useful standard for global urban characterisation, class-based approaches may overlook important local differences, especially in cities like those in Australia, where urban forms and functions can differ substantially from the definitions underlying global LCZ typologies[33,34]. New datasets and modelling platforms, such as digital twin cities, are becoming increasingly available that describe the three-dimensional structure of buildings and even trees at high resolutions. These datasets can be used to determine local characteristics without relying on classes, but their direct use in numerical weather and climate modelling requires post-processing to conform to the required inputs of climate models[32].

Furthermore, there is a significant gap in data concerning future urban development. Current datasets do not adequately capture how urban areas are expected to evolve, ignoring future expansions, densification trends, changes in land use, or the influence of planning policies such as zoning regulations and greenfield developments. With Australia's population projected to grow substantially over the coming decades, integrating future scenarios based on population projections and plausible socioeconomic pathways (such as SSPs) is essential for climate modelling that reflects realistic urban futures. Without forward-looking data, we are unable to accurately assess the climate and weather impacts of urban growth or evaluate the implications of different development strategies. Moreover, we lack the tools to guide or limit urban expansion into areas highly exposed to climate hazards, such as floodplains, coastal estuaries and lagoons, the peri-urban zone or heat-prone regions.

To fully understand and model the impacts of urban environments, it is necessary to account for how human activities contribute to energy production and heat emissions. This aspect is essential for evaluating the effectiveness of strategies aimed at reducing carbon emissions and achieving net-zero targets, as well as for understanding how these factors influence weather and climate across different scales. Currently, consistent data on these factors is not adequately represented in our climate models. Closing this gap directly supports several of 2025 Australia's National Science and Research Priorities, notably *Transitioning to a net zero future*, *Supporting healthy and thriving communities*, and *Building a secure and resilient nation*[35].

Finally, it is not enough to simply create high-resolution urban datasets. For these datasets to be usable by climate modellers and researchers, they must be curated, standardised, and delivered in formats that align with existing climate modelling frameworks. Just as importantly, they need to be co-located with high-performance computing (HPC) environments, such as the National Computational Infrastructure (NCI) supporting nationally-relevant products from climate simulations (e.g. Coupled Model Intercomparison Project Phase 6 (CMIP) and NARCliM[36]). Without this, their potential for integration into land surface models and climate projections remains unrealised.

> **Gap**: Lack of nationally consistent input datasets for cities that represent the *current* spatial variability in urban form, fabric, and function. These datasets should be findable, accessible, interoperable with other systems, reusable, and available under an open license (following the FAIR data principle). While some datasets exist, they are often not publicly available and may require commercial subscriptions that are cost-prohibitive for research, model development, and downstream applications.
>
> **Gap:** Lack of nationally consistent input datasets for cities, representing *future scenarios* of urban expansions and densifications.
>
> ***Recommendations #1:*** *Leverage existing very high-resolution, building-resolving datasets (e.g., satellite, LiDAR, and AI-based products at 2–100 m resolution) to generate national datasets of urban characteristics, including land cover, building heights, and morphology information needed for urban climate modelling.*
>
> ***Recommendation #2:*** *Develop nationally consistent projections of urban expansion and densification, based on existing zoning laws, planned development, population growth, as well as multiple socioeconomic pathway levels (e.g., SSPs), to be integrated into climate projections.*

> **Gap**: Lack of nationally developed input datasets that represent (current and future) spatial variability and changes in anthropogenic waste heat flux and energy production in cities.
>
> ***Recommendation #3:*** *Develop nationally consistent datasets of anthropogenic heat and energy production with net-zero ambitions for current and future scenarios, aligned with projections of urban expansion (Recommendation #2), that can be integrated into climate projections.*

> **Gap**: Lack of coordinated processes and standards[37] to ensure that urban datasets (once developed) are curated, documented, regularly updated, and made available in formats suitable for ingestion into land surface and climate models. Furthermore, these datasets are not systematically co-located with HPC resources, limiting their utility for national-scale modelling efforts.
>
> ***Recommendation #4****: Develop a national framework for urban climate data curation and integration, including standardisation protocols, metadata documentation, update mechanisms, and co-location with HPC environments. Ensure that datasets are provided in formats compatible with land surface models and regional/global climate simulations.*

### b. Integrated urban climate observation networks

Assessing the accuracy of urban models for the Australian context relies on systematic and comprehensive observational networks at scales relevant to urban-resolving applications. This is a key area where Australia currently lacks nationally consistent urban climate data and capability, reflected both in the absence of long-term investment in urban observational infrastructure and in limited national expertise in designing and operating such long-term urban networks. Without targeted observations, our ability to evaluate models, understand key processes, and support evidence-based urban climate adaptation is severely constrained.

The gap is most apparent for *in-canopy measurements* (below building height, both fixed and spatial transects) due to the lack of integrated and coordinated measurement campaigns. Traditional synoptic-scale weather observing networks follow international guidelines that require minimising the influence of buildings and artificial surfaces. The Australian Bureau of Meteorology, the government agency tasked with national weather observations, therefore places its "urban" observing stations in large parks or grassy fields within airports rather than in places where people live. While this approach is suitable for large-scale weather monitoring, it makes these networks ill-suited for capturing local-scale urban conditions[38]. Some observational networks have been established with the intent to capture the diversity of urban microclimates[39–41], but these efforts lack national coverage, are often fragmented with varying levels of accessibility, do not follow consistent methodologies, and are rarely maintained over extended periods. To date, Australia lacks a single consolidated platform that provides information, let alone access, to the full range of microclimate observations collected by private, academic, government, and community sectors, leaving substantial investments of time, funding, and effort underutilised.

*Flux tower observations*, which measure exchanges of energy, moisture, and momentum between the surface and atmosphere, are particularly limited for Australian cities. Only one campaign was conducted in Melbourne for a short duration (one-year measurement in 2004), and no sustained national effort has followed[42]. Existing international flux tower datasets also poorly represent Australia's climatic conditions (Figure 1). The climatological coverage of global urban flux and meteorological tower sites (black dots) used for model evaluation[43] fails to capture the diversity of Australia's major cities (orange dots) when compared with more than 70,000 urban areas worldwide (grey dots). This gap limits our ability to understand and model land–atmosphere interactions in Australian urban environments.

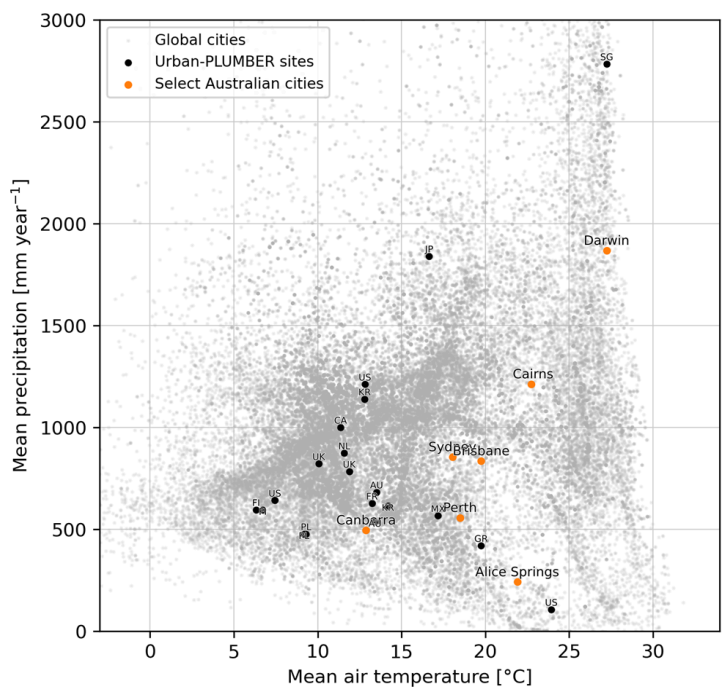

**Figure 1.** Climatology of urban flux tower sites (black circles) compared with Australian major cities (orange circles) and 70,000 global urban areas (grey dots) adapted from[43]. Only one of the 20 urban flux towers (indicated as AU) was sited in Australia[44].

Beyond flux towers, boundary-layer, multi-variate, and spatial transect observations are also critically lacking. These include vertical profiles of wind, turbulence, aerosols, and thermodynamic variables[45], as well as surface radiation budgets and air quality. Comprehensive urban observation networks rely on a diverse array of instruments, including Doppler wind lidars (DWL), microwave profilers, radiosondes, and sun photometers. Each provides distinct but complementary insights into urban boundary-layer dynamics and land–atmosphere exchanges[46–48]. These measurements are essential for evaluating model physics, closing surface-atmosphere energy budgets, and supporting applications beyond climate modelling, such as urban health assessments or infrastructure planning for major events like the 2032 Brisbane Olympics.

In contrast, international programs provide useful precedents. For example, the **PANAME** observation campaign in Paris (https://paname.aeris-data.fr) is an extensive, multi-method effort that combines remote sensing, flux towers, surface meteorology, and in-canopy measurements to assess urban processes and evaluate models[47,48]. Similarly, in Berlin, the **Urban Climate Observatory (UCO) Berlin**, a core part of the Urbisphere project shown in Fig. 2, provides long-term multivariate observations from the urban canopy to the boundary layer, including eddy covariance flux measurements, radiation monitoring, Doppler wind LiDAR, microwave profiling, and crowdsourced environmental data[46]. This multi-scale, multi-variable system exemplifies the kind of sustained and coordinated observation network needed to support both fundamental urban climate science and applied modelling initiatives.

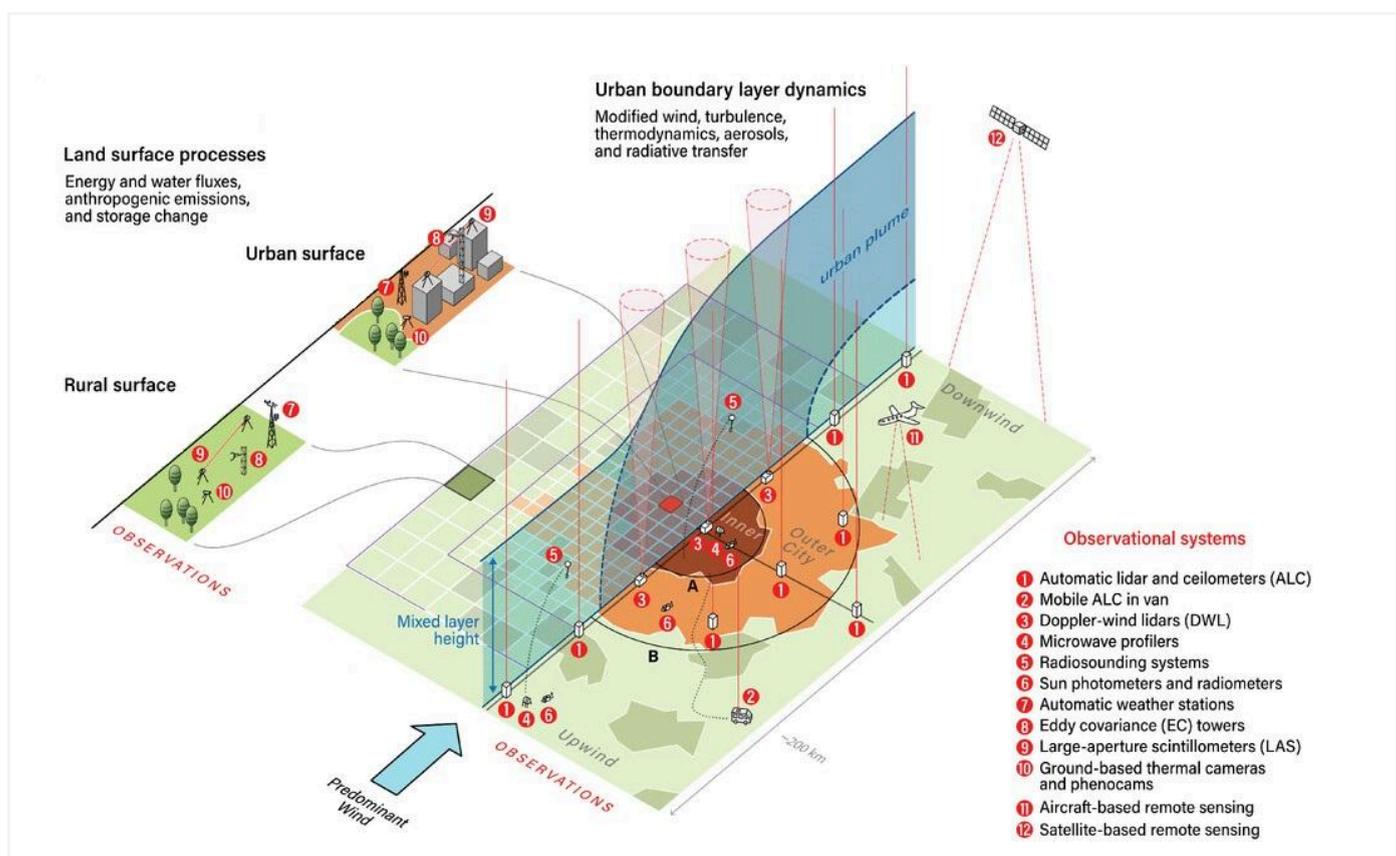

**Figure 2.** Modified from Fenner et al.[46], this schematic illustrates the comprehensive suite of observational methods used in the Urbisphere project in Berlin to characterise urban land surface processes and boundary layer dynamics. It highlights the spatial distribution and layering of instrumentation across rural-to-urban transects, including flux towers, lidar systems, radiosondes, satellite and aircraft-based sensors, and surface monitoring platforms, demonstrating the multi-scale, multi-variable approach required for robust urban climate observation.

In addition to addressing gaps in conventional observations, there is also a pressing need to explore and leverage non-traditional or unconventional data sources that, while possibly less standardised, can still offer significant spatial and temporal insights. These include emerging opportunities such as hyperspectral and thermal infrared data from geostationary satellites (e.g., Himawari[49]), anonymised mobile phone data that reflect population dynamics and heat exposure patterns[50,51], and citizen science observations (e.g., participatory temperature or air quality monitoring[6,52]). While these sources may come with calibration and consistency challenges, they can complement traditional networks by offering broader spatial coverage or continuous high-frequency data. Integrating such data streams will be essential for characterising diverse urban climate regimes and improving the robustness of urban climate model evaluation.

> **Gap**: Sparse, inconsistent, or short duration of in-canopy (near-surface) urban observations (focused on variables such as temperature, air quality, and water within the urban canopy relevant to people and infrastructures) across urban areas in Australia.
>
> **Gap**: Very limited flux tower measurements over urban areas in Australia, with the last one concluded in 2004. Similarly, limited boundary layer and surface measurements over cities, with most existing weather measurements sited in locations such as parks or airports, failing to capture representative urban microclimates.
>
> ***Recommendations #4***: *Develop a nationally consistent, high-quality, and long-term funded observation network for urban areas, including measurements both within and above the urban canopy. This should include variables such as radiation budgets, energy and carbon fluxes, and meteorological profiles to support robust model evaluation and climate research.*
>
> ***Recommendation #5***: *Identify and evaluate the potential of unconventional and emerging data sources (e.g., hyperspectral satellite data, mobile phone-derived observations, and citizen weather stations) for improving spatial and temporal coverage of urban climate observations, particularly in areas lacking dense instrumentation.*

> **Gap**: Lack of integrated platforms to consolidate urban data collected by state/local Governments and university projects.
>
> ***Recommendations #6***: *Prioritise the coordination, standardisation, and visibility of existing urban climate data collected by disparate groups. Applying FAIR (Findable, Accessible, Interoperable, and Reusable) data principles[53] will maximise the value of existing datasets and support broader uptake by climate modellers, planners, and decision-makers.*
>
> ***Recommendation #7***: *Establish and integrate data platforms that integrate datasets from national urban datasets (such as AURIN) with local- and state-level observation networks, together with modelling outputs incorporated in ACCESS-NRI. This ensures that Australia's growing volume of research data is in accordance with the National Digital Research Infrastructure Strategy[54].*

## II.B. Developing and assessing fit-for-purpose models

### a. Establishing and coordinating a national urban modelling capability

Numerous processes and submodels are required to capture the complexity of urban environments. These range from detailed representations of anthropogenic heat and emission sources to the influence of urban geometry, vegetation, and water networks. Accurate modelling of these processes affects simulated temperatures, air quality, wind patterns, precipitation, and other climate phenomena in cities[55,56], with impacts evident even at the global scale[12,57]. Without appropriate parameterisation, urban schemes produce unreliable predictions and projections, leading to inaccurate assessments of urban impacts on weather and climate.

The rationale for including urban representation in climate models varies with the intended application. For most global and regional-scale studies, the priority is to improve representation of land-atmosphere interactions above the canopy (roof height), to inform large-scale atmospheric simulations[58]. In contrast, applications focused on climate impacts and risks within cities emphasise processes below roof height, inside the urban canopy, where accurate predictions and projections of temperature, heat stress, air quality, rainfall, and energy use directly influence health systems, infrastructure, and community resilience[59,60]. These different applications require different modelling approaches. Having the appropriate physics is essential when addressing 'what if' questions about adaptation and mitigation, as models with incorrect process representations can lead to incorrect conclusions. Both perspectives require robust urban modelling, but the definition of appropriate in terms of physics, inputs, and resolution depends on the specific scale and intended application.

A key gap in Australia's current modelling capability is the inconsistent or absent representation of urban areas across scales. At the global scale, the problem begins with how cities are depicted in coupled climate models. Major Australian cities and towns can be classified as ocean grid points in over 70% of CMIP6 models (Fig 3). This means that if we extract future climate projections at the grid points corresponding to our major urban areas, even when using multi-model ensemble averages, the results are based on ocean-atmosphere physics rather than processes relevant to cities. This underrepresentation stems partly from land-sea masks defined primarily by ocean modellers and from coarse spatial resolution, which excludes the influence of urban areas. However, this is not purely a resolution problem. CMIP6 models with grid sizes of 50 to 100 km perform similarly to those with around 250 km resolution, with coastal cities poorly represented in both cases. While higher resolution can improve results, without appropriate urban parameterisations, it may also amplify errors.

Even when land is correctly identified, most global models used in CMIP6 do not include any urban physics. This is gradually changing as more global models introduce urban representation and physics[61–64]. However, CABLE, the LSM used in Australia's global climate projections, currently lacks any urban representation. As global projections increasingly operate at higher resolutions, it is critical that Australian LSMs also include urban representation so they can accurately simulate modifications to surface fluxes over cities.

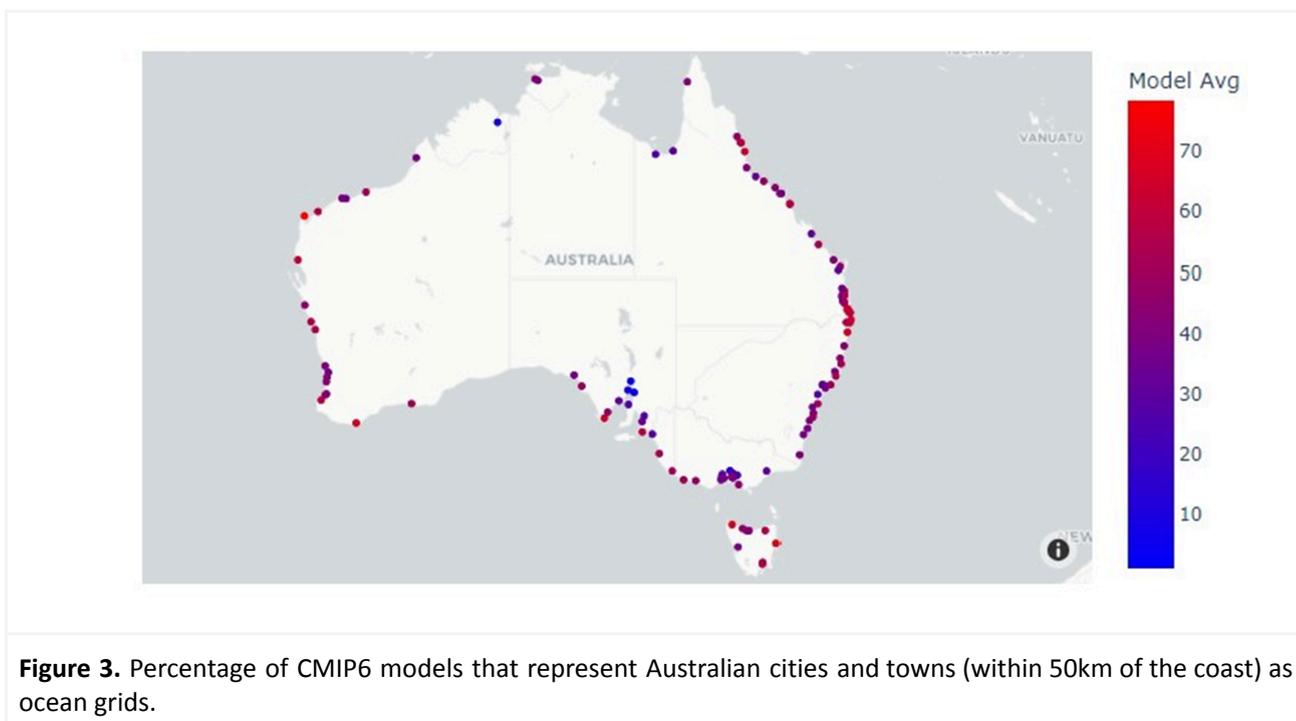

**Figure 3.** Percentage of CMIP6 models that represent Australian cities and towns (within 50km of the coast) as ocean grids.

At the regional scale, urban representation becomes even more critical for producing actionable climate information for cities, and as a result, it has received greater attention. However, regional modelling for Australian cities still faces major constraints. Different regions use different regional climate models (RCMs), most of which were developed for the Northern Hemisphere and lack urban parameterisations that capture Australia-specific climate drivers and phenomena. Urban processes such as vegetation, hydrology, and chemistry, and the interaction of dense built environments with local hydrology and air quality, are often simplified or excluded, and in-canopy variables like temperature, humidity, and air quality are rarely simulated at the resolution needed for impact assessments. In addition, descriptive datasets of urban form, fabric, and function are inconsistent or unavailable nationally, and the resolution of projections often remains too coarse to meet emerging stakeholder needs, which can require kilometre or sub-kilometre detail.

Some initiatives have made progress. For example, NARCliM[36] has enabled single-layer urban canopy schemes within multi-model downscaling frameworks over south-eastern Australia, providing a key step towards incorporating urban processes into projections that could be extended to include more realistic urban representations. However, the level of detail required to capture within-canopy variability and the full complexity of surface-atmosphere interactions across Australia's diverse urban environments remains absent from most modelling efforts.

Urban landscapes can no longer be ignored (nor should they be considered as rocks with no vegetation). Their representation needs to be scale-dependent and process-based, suitable for a hierarchy of model prediction and projection systems across (spatial and temporal) scales. Achieving this requires an accurate depiction of the urban landscape (Section II.A) and models with the capability to ingest, process, and fully utilise these datasets. Building such capability means integrating accurate, high-resolution datasets describing the current and future urban form, fabric, and function, and ensuring that projections are produced at resolutions that meet the emerging stakeholder demand for actionable climate information in cities. Comparable international efforts, such as the Global Energy and Water Exchanges (GEWEX) program Cloud System Study[65] or current WCRP CORDEX initiatives, demonstrate the value of bringing observations, process models, parameterisation development, and community coordination together into one strategy. A similar integrated approach is now required for Australia's urban climate capability.

**Gap**: Major Australian cities and towns are represented as ocean grids by over 70% of CMIP6 models (Figure 3).

**Gap**: Global climate projections produced by Australian models (as used in Coupled Model Intercomparison Project (CMIP) experiments) have no representation of urban areas. This is primarily due to the lack of any urban representation in the Community Atmosphere Biosphere Land Exchange (CABLE).

*Recommendation #8:* *As global projections reach higher resolutions, enable urban representation in Australian global projections that accurately represent surface flux modifications over cities.*

**Gap:** Inconsistencies in regional modelling frameworks, including reliance on tools designed for the Northern Hemisphere that overlook Australian-relevant climate drivers and phenomena.

**Gap:** Inconsistencies and limitations in regional climate projections for cities, particularly in the way urban processes (e.g. form, vegetation, hydrology, chemistry) are parameterised in land surface models and in the representation of in-canopy variables such as temperature, humidity, and air quality.

**Gap:** Lack of or limited representation of future urban expansion and densification, as well as city-scale net-zero ambitions in climate projection scenarios.

**Gap:** Limited ability to capture extremes (beyond mean states) in urban areas, due to incomplete process representation and interactions with 3D urban form, constraining assessment of risks to health, infrastructure, and insurance.

**Gap:** Insufficient nationally consistent, fine-resolution (km-scale) climate projections that cover all major Australian cities over long time horizons (including end-of-century), produced through multi-model ensembles and downscaling approaches tailored to Australian climate drivers.

*Recommendation #9:* *At the regional scale, provide nationally consistent, Australia-wide, km-scale, and long-term climate projections informed by multi-model downscaling products, current and future urban datasets, and benchmarked urban canopy representations. (informed by process understanding?)*

**Gap**: CABLE, an LSM supported by ACCESS-NRI and used for climate modelling over Australia by research communities and for CMIP experiments, does not have an urban parameterisation.

*Recommendation #10:* *Develop the modelling capability to represent urban datasets and processes in the CABLE LSM, such that we quantify a) the impact of cities on surface fluxes (prioritised at the global scale), and b) the impact of climate projections on in-canopy variables (prioritised at the regional scale).*

*Recommendation #11:* *Develop scale-dependent urban parameterisations suitable for a hierarchy of global to regional models. This should include the inherent capability to ingest and make use of new, high-resolution urban datasets through model-data fusion and parameter estimation, without requiring major redevelopment of the modelling framework.*

### b. Model comparison, evaluations, and benchmarking

Each model is developed with a specific application in mind and assessed using observation of limited climatic and urban conditions. Ensuring that urban climate modelling is ready to answer unique challenges in Australian cities requires consistent evaluation based on local use cases and city characteristics. This relies on systematic and comprehensive observational networks at scales relevant to urban-resolving applications.

At present, our ability to evaluate modelling capabilities for Australian cities is limited by both the lack of suitable observational data (Section II.A) and the absence of coordinated intercomparison projects that systematically assess process-based urban parameterisations across scales. This is part of a broader global issue, as most international evaluation efforts have focused on fluxes above the urban canopy layer, neglecting critical in-canopy variables such as temperature, humidity, and air quality. In addition, evaluations of historic climate projections often rely on reanalysis products like ERA5 that omit urban representation entirely, reducing their accuracy for city-scale assessments. Reanalysis products that incorporate urban representation through modelling and data assimilation (such as the Bureau of Meteorology's convective-permitting BARRA-C2[66]) offer a more robust basis for evaluating urban climate projections and should be prioritised for further development and application.

> **Gap**: Lack of intercomparison projects that focus on in-canopy variables in addition to land surface fluxes to the atmosphere. Previous international inter-model comparison projects have primarily focused on model evaluation based on surface fluxes above the urban canopy layer [58].
>
> ***Recommendation #12:*** *Develop common evaluation frameworks for intercomparison and benchmarking of urban models and their performance in climate projections. This should be informed by nationally consistent urban datasets for urban canopy parameters as well as surface fluxes.*
>
> **Gap**: Inaccurate evaluations of climate projections in the historic period based on reanalysis data that lack urban representation (such as ERA5).
>
> ***Recommendation #13:*** *Develop and use reanalysis products that include urban representations (through both modelling and data assimilation) for the evaluation of regional climate projections over cities. An example of reanalysis products with urban representation in modelling includes the convective-permitting Bureau's Atmospheric Regional Reanalysis for Australia (BARRA-C2)[66], which should be further evaluated and benchmarked for Australian cities.*

## II.C Building an active community to advance Australia's understanding of weather-climate interactions in cities

Developing and maintaining an advanced modelling capability requires an active and connected network of multidisciplinary experts and users. The research itself must be a community effort, drawing together expertise from global and regional climate science, remote sensing, air quality, hydrology, computational fluid dynamics, and other fields. Sustained engagement with policymakers, land use planners, architects, and other stakeholders is then essential to drive innovation and ensure that models remain accurate and relevant. This community must be supported to enable cutting-edge knowledge exchange that addresses the evolving needs of Australia's cities and communities.

Alongside this community-building effort, key emerging scientific questions need to be addressed in the context of Australian cities. There is a need to better quantify the influence of Australian cities on regional weather and climate extremes (including high-impact weather events such as short-duration rain and hail), and to identify the climate thresholds that drive vulnerabilities in urban systems. Progress is currently constrained by limited convective-permitting, high-resolution (1 km-scale) climate projections that include urban representation, which are essential for simulating key processes such as sea breezes, foehn winds, heatwaves, and extreme rainfall. Similarly, coordinated research is needed to assess the role of AI/ML in enhancing or complementing physical models, addressing large data needs, and improving urban climate projections.

Strengthening the national community and capability in these areas will also provide cities with better tools to test climate change adaptation strategies. As more frequent and intense heatwaves, rainfall, and flooding events are observed and projected, cities urgently need robust ways to evaluate the effectiveness of proposed interventions under a range of future climate scenarios, as well as future urban expansion/densification pathways, such as greenfield developments and conversion of low-rise low-density commonly seen in Sydney[34].

This effort is also essential in light of Australia's shift to **mandatory climate-related financial disclosure**[67]. As organisations are now required to report physical climate and transition risks, there will be increasing demand for more granular, localised climate risk information, well beyond what can be provided by global or even standard regional climate models. While disclosure frameworks rely on a suite of tools, understanding how urban characteristics amplify or reduce climate-related risks (and create opportunities for innovation, adaptation, and resilience) is best supported by modelling approaches designed with cities in mind. Without this, organisations (and the communities they serve) remain exposed to blind spots in risk assessment.

AI/ML may offer the potential to accelerate model development, reduce computational costs, and unlock new insights from large datasets. However, its application in urban climate modelling remains underexplored. This limits confidence in its use for urban-relevant processes such as surface energy balance, air temperature, and anthropogenic heat release. ML methods are often trained on high-resolution datasets (e.g., Large Eddy Simulation or convection-permitting simulations) to inform parameterisations for coarser-resolution models[68]. However, the extent to which these schemes generalise across urban forms, climates, and spatial scales is still unclear. Without robust frameworks for validation and uncertainty quantification, there is a risk of embedding opaque or poorly constrained ML-derived components into models used for infrastructure and climate risk planning. This gap has practical consequences for infrastructure design, site-specific risk assessment, and the strategic placement of observational networks. It also hampers efforts to integrate ML workflows into national-scale modelling infrastructure or use ML-informed downscaling in climate disclosure and resilience assessments.

From an application perspective, the urban climate community also lacks clarity on the data requirements for ML/AI approaches, where these data should be stored, and how they can be made computationally accessible for training. Effective deployment will require co-location of large datasets (e.g., multi-decadal high-resolution climate simulations or national-scale remote sensing) with GPU-enabled HPC. Without clear frameworks for data governance and access, ML/AI use in urban climate projections will remain limited. Given the resource intensity of large-scale training, ensuring efficient use of national infrastructure and renewable-powered facilities will also be important.

> **Gap**: Lack of a consolidated national scientific community dedicated to the development and exchange of knowledge on urban representation in climate modelling and projections.
>
> *Recommendation #14: Build a community of practice as well as infrastructure support for urban climate research in Australia. This forms effective dialogues and best practices for incorporating urban processes (including hydrology and chemistry) into multi-scale, multi-model climate analyses*
>
> **Gap**: Lack of national infrastructure support that integrates, consolidates, and supports existing scientific understanding, datasets, and modelling methods for accurate representations of urban areas in climate projections.
>
> **Gap**: There was no scope in the current ACCESS-NRI work plan to provide infrastructure support, documentation, and guideline development in the urban model development.
>
> *Recommendation #15: Integrate and streamline urban representation into the ACCESS-NRI capabilities and develop the knowledge base that can further support diverse communities.*
>
> *Recommendation #16: Engage with global initiatives such as the WCRP CORDEX flagship study on URBan environments and Regional Climate Change (URB-RCC) to incorporate global best practices in regional climate projections over cities.*

*Gap*: Limited convective-permitting regional climate projections at high resolutions (km-scale) with urban representations, limiting the quantification of cities' impact on regional climates.

*Gap*: Limited attribution analyses on the impact of cities on regional weather and climate extremes in Australia (due to a lack of model capabilities) that further limit process-based, scale-dependent urban parameterisation in climate projections.

*Recommendation #17:* *Prioritise urban representation in convective-permitting (km-scale) regional climate projections as the key tool for quantifying the cities' two-way interaction with regional processes (such as sea breeze impact) and extremes (such as precipitation and heatwaves), informing the development of fit-for-purpose parameterisations that can accurately represent them in climate projections.*

*Gap*: Limited understanding of the role of AI/ML methods in a) downscaling climate projections in cities and b) representing urban physics in LSMs. Key uncertainties include the choice and suitability of training datasets (compounded by limitations in urban observations), the interpretability of model outputs, the robustness of results under future climate conditions, and the integration of ML-derived outputs into existing modelling frameworks.

*Recommendation #18:* *Parallel to scientific knowledge focused on the physical modelling of urban impacts, further develop our scientific understanding of the advantages and limitations of ML in a) downscaling climate projections in cities and b) representing urban physics in LSMs. This should include:*

- *Identifying priority problem domains where ML/AI offers clear advantages, such as statistical downscaling using multi-source observations, derivation of scale-aware parameterisations from high-resolution model data, and automated parameter optimisation.*
- *Establishing benchmark datasets and evaluation frameworks to assess ML/AI performance against physical models and observations across a representative range of Australian urban environments.*
- *Ensuring data storage and access arrangements that support computationally intensive ML/AI training, including co-location of data and compute resources on national HPC facilities.*
- *Developing workflows that integrate ML/AI outputs into the existing urban climate modelling chain in a way that is transparent, reproducible, and adaptable to future data sources.*

## III.  Conclusions: From Vision to Action

Many of the recommendations provided here echo those of the ATSE report nearly 15 years ago[21]. Modelling remains our primary tool for understanding how different strategies for urban landscape design will influence extremes of weather and climate. To make these models relevant to Australian cities, urban-scale processes must be systematically integrated into national modelling capabilities, particularly the Australian Community Climate & Earth System Simulator (ACCESS). This requires targeted research to develop a hierarchy of models that are fit-for-purpose for a range of urban applications, from human-scale analyses to global assessments. A systematic observation program is equally critical, incorporating meteorological, flux, and remote sensing data from cities with diverse climates and characteristics. In parallel, emerging high-resolution, building-resolving datasets describing urban land cover, building height, and morphology should be harnessed to create a publicly available national map of urban characteristics. Such advances will enable nationally consistent climate projections for cities that account for urban expansion and densification, and allow us to address key fundamental questions about the interaction between urban form and climate.

These four areas (data, observations, models, community) are not separate silos but interdependent pillars that only have impact when pursued together. Their integration into a single, **nationally coordinated science strategy** is essential if Australia is to move beyond piecemeal progress. This integration provides the foundation for a coherent national program and positions Australia to lead internationally in urban climate research, with lessons applicable well beyond our region.

The experience following the ATSE report underlines a hard truth: without a dedicated, funded body to lead and coordinate research and practice, even well-defined recommendations remain unfulfilled. Despite identifying many of the same gaps we face today, progress stalled due to fragmented efforts, lack of critical funding, and absence of a unified implementation plan. To avoid repeating this pattern, we must establish a specifically funded and coordinated program to support the national science strategy, embedded within and working alongside national research infrastructures such as ACCESS-NRI and AURIN.

The cost of inaction will be high: missed opportunities to reduce heat stress and improve liveability, higher infrastructure and health costs from poorly adapted urban design, inefficient use of water and energy, and greater economic losses from climate extremes. In contrast, investment in fine-grained urban climate modelling will deliver significant returns. With the right training and tools, planners and engineers can use modelling outputs to guide more efficient infrastructure planning, while targeted cooling strategies can reduce peak energy demand, and communities can be supported to adapt even in existing built-up areas. Better modelling also enables more resilient water management in and around cities, where supply, stormwater, and demand pressures intersect most strongly.

Coordinated national action will not only reduce risks but also create opportunities for innovation and growth. Investment in this capability will strengthen Australia's position in green infrastructure, energy systems, building design, and climate-responsive planning, turning better science into both avoided costs and new business opportunities. Coordinated national action is therefore not just a scientific necessity but a strategic investment in Australia's economic, environmental, and societal resilience, and an opportunity for Australia to provide global leadership in how urban climate science informs adaptation and risk management.

Finally, Australia's diverse climates, from tropical to desert to alpine, make it a natural laboratory for developing urban climate modelling approaches with global applicability. Lessons learned here can inform strategies in regions facing similar extremes globally.